\documentclass[aps,prd,12pt,preprint,eqsecnum,tightenlines,
amsfonts,amssymb,amsmath,byrevtex,showpacs]{revtex4}
\begin{document}
\newcommand{\dR}{\mathbb R}
\newcommand{\dC}{\mathbb C}
\newcommand{\dS}{\mathbb S}
\newcommand{\dZ}{\mathbb Z}
\newcommand{\aff}{\mathfrak{aff}}
\newcommand{\Aff}{\mathfrak{Aff}}

\title{Quantization and spacetime topology}

\author{W\l odzimierz Piechocki}
\affiliation{So\l tan Institute for Nuclear Studies, Ho\.{z}a 69, 00-681 
Warszawa, Poland\\e-mail: piech@fuw.edu.pl \\}
\date{\today}
\begin{abstract}
We consider classical and quantum dynamics of a free particle in de Sitter's 
space-times with different topologies to see what happens to space-time  
singularities of removable type in quantum theory. We find analytic 
solution of the classical dynamics. The quantum dynamics is solved by 
finding an essentially self-adjoint representation of the algebra of 
observables integrable to the unitary representations of the 
symmetry group of each considered gravitational system. The dynamics of a 
massless particle is obtained in the zero-mass limit of the massive case. 
Our results indicate that taking account of global properties of space-time 
enables quantization of particle dynamics in all considered cases.

\end{abstract}
\maketitle

\section{Introduction}

Cosmological data indicate that the Universe expands, so in the past 
it could be in a special state characterized by physical fields with 
extremally high densities. There are also theoretical indications that our 
universe emerged from a very special state: the well known class of solutions 
to the Einstein equations called the FLRW universes suggest that in the past 
our universe could be in a state with blowing up Riemann tensor components 
or scalar curvature and with blowing up energy density. There is a common 
believe that to analyse this state properly one should include quantum effects.
The struggle for quantum gravity lasts about 70 years. There is a real 
progress, but we believe that one should first understand the nature of 
space-time singularities in a quantum context. 
This understanding may mean changing some of the principles underlying 
quantum mechanics or general relativity. The insight into the problem may be 
achieved by studying some suitable toy models which include both space-time 
singularities and quantum rules. In what follows we present results 
concerning one of such models which is quantization of dynamics of a test 
particle in  singular and corresponding regular space-times.

Recently it was found \cite{GW1,GW2,GW3,GW4} that classical and quantum 
dynamics of a particle in a curved space-time seems to be  sensitive 
to the topology of space-time. Our aim  is  examination of this dependence 
in details. It enables the understanding of the nature of removable type 
singularities of space-time.
 
We use  the group theory oriented quantization (GTQ) scheme,  which 
we have already applied to  simple gravitational systems 
\cite{GW1,GW2,GW3,GW4,GW5}. Our method is similar to 
the GTQ method initiated by Isham \cite{CJI} and Kirillov \cite{AK1}.

In what follows we examine classical and quantum dynamics of a particle in 
two-dimensional space-times with different topologies. 
We  carry out all calculations rigorously which enables 
complete discussion of considered problems. 
In the last section  we make the argument that our results can be extended to 
higher dimensions. 

In Sec. II we present the dynamics of a particle in regular 
space-time. Application of the standard GTQ method leads to well defined 
results. 

Analyses of  particle dynamics in singular space-time is carried out in 
Sec. III. The GTQ method  needs some modification to be applicable in this 
case, since the relation between local and global properties of considered 
system cannot be directly modeled by mathematics connecting  Lie group 
and its Lie algebra (consequently also at the level of representations). 
However, redefinition (for the purpose of quantization) of the notion of 
local symmetries of a gravitational system enables the quantization. 
The problem of quantization in this case is directly connected 
with the problem of space-time singularities. We present the solution in 
case space-time has singularities of removable type.

Secs. II and III deal with a particle with a non-zero mass. 

In Sec. IV we present the dynamics of a massless particle. It is obtained 
in the zero-mass limit from the massive particle dynamics. 

We conclude in Sec. V. The last section comprises the list of references. 

\section{Particle on hyperboloid}

The considered space-times,  $V_p$ and $V_h$, are of de Sitter's type. 
They are defined to be \cite{GW5}
\begin{equation}
V_p = (\dR\times \dR,~\hat{g})~~~~~~\text{and}~~~~~~~~
V_h = (\dR\times \dS,~\hat{g}).
\end{equation}
In both cases  the metric $g_{\mu\nu} :=(\hat{g})_{\mu\nu} 
~~(\mu ,\nu= 0,1)$ 
is defined by the line-element
\begin{equation}
ds^2 = dt^2 - exp(2t/r)~dx^2,
\end{equation}
where $r$ is a positive real constant.

\noindent
It is clear that (2.1) includes all possible topologies of de Sitter's 
type space-times in two dimensions which makes our examination complete.
$V_p$ is a plane with global $(t,x)\in \dR^2$ 
coordinates. $V_h$ is defined to be a one-sheet 
hyperboloid embedded in 3d Minkowski space. There exists an isometric 
immersion map \cite{SHE} of $V_p$ into $V_h$
\begin{equation}
 V_{p} \ni (t,x)\longrightarrow (y^0,y^1,y^2) \in V_h,
 \end{equation}
where
$$ y^0:=r\sinh (t/r) +\frac{x^2}{2r}~\exp(t/r),~~~
y^1:=-r\cosh (t/r)+\frac{x^2}{2r}~\exp(t/r),~~~
 y^2:= -x \exp(t/r),$$
and where 
\begin{equation}
 (y^2)^2+(y^1)^2-(y^0)^2=r^2.
\end{equation}
Eq. (2.3) defines a map of $V_p$ onto a simply connected 
non-compact half of $V_h$. Thus, $V_p$ is just a part of $V_h$. 
One can check that the induced metric on $V_h$ coincides with 
the metric defined by (2.2).

It is known \cite{SHE} that $V_p$ is geodesically incomplete. However, all 
incomplete geodesics in $V_p$ can be extended to complete ones in $V_h$, 
i.e. $V_p$ has removable type singularities. $V_p$ and $V_h$ are 
the simplest examples of space-times with constant curvatures and with 
noncompact and compact spaces, respectively. 

An action integral, $\mathcal{A}$, describing a free relativistic particle 
of mass $m$  in gravitational field  $g_{\mu \nu}$ is proportional to the 
length of a particle world-line and is given by
\begin{equation}
\mathcal{A}=\int_{\tau_1}^{\tau_2}~L(\tau)~d\tau,~~~~~L(\tau)
:=-m\sqrt{g_{\mu\nu}(x^0(\tau),x^1(\tau))\;\dot{x}^\mu (\tau)
\dot{x}^\nu (\tau)},
\end{equation}
where $\tau$ is an evolution parameter, $x^\mu$ are 
space-time coordinates and $\dot{x}^\mu := dx^\mu/d\tau $. It is assumed 
that $ \dot{x}^0 >0 $, i.e., $x^0$ has interpretation of time monotonically 
increasing with $\tau$. 

The Lagrangian  (2.5) is invariant under the reparametrization
$ \tau\rightarrow f(\tau)$.  
This gauge symmetry leads to the constraint 
\begin{equation}
 G:= g^{\mu\nu}p_\mu p_\nu -m^2=0,
\end{equation} 
where $g^{\mu\nu}$ is the inverse of $g_{\mu\nu}$ and 
$p_\mu := \partial L/\partial\dot{x}^\mu$ are canonical momenta. 

Since we assume that a free particle does not modify the geometry of 
space-time, the local symmetry of the system is defined by the set 
of all Killing vector fields of  space-time (which is also the local symmetry 
of the Lagrangian $L$). The corresponding dynamical integrals have the 
form \cite{HHS}
\begin{equation} 
D=p_\mu X^\mu,~~~~~\mu= 0,1 , 
\end{equation}
where $X^\mu$ is a Killing vector field.

\noindent 
The physical phase-space $\Gamma$ is defined to be the space of all particle 
trajectories \cite{JJS} consistent with the dynamics of a particle and with 
the constraint (2.6).

\subsection{Classical dynamics}

Since we consider the dynamics of a material particle (i.e. moving along 
timelike geodesics) on the hyperboloid (2.4), the symmetry group of $ V_h $ 
system is the proper orthochronous Lorentz group $SO_0(1,2)$.  

If we parametrize (2.4) as follows 
\begin{equation}
y^0=-\frac{r\cos \rho /r}{\sin \rho /r},~~~~y^1=\frac{r\cos \theta /r}
{\sin \rho /r},~~~~y^2=\frac{r\sin \theta /r}{\sin \rho /r},
\end{equation} 
where $0<\rho < \pi r,~0\leq \theta < 2\pi r$,
the line-element on the hyperboloid (2.4) reads 
\begin{equation}
ds^2 = (d\rho^2 - d\theta^2)\sin^{-2}(\rho/r),
\end{equation}
and the Lagrangian (2.5) is given by 
\begin{equation}
L= -m~\sqrt{\frac{\dot{\rho}^2 - \dot{\theta}^2}{\sin^2  (\rho /r)}} .
\end{equation}
Since we consider only timelike trajectories ($|\dot{\rho}|>|\dot{\theta}|$), 
the Lagrangian (2.10) is well defined.

The infinitesimal transformations of $SO_0(1,2)$ group (rotation 
and two boosts) have the form    

\begin{eqnarray}
(\rho,~\theta)\longrightarrow(\rho,~\theta+a_0 r),~~~~~~~~~~~~~~~~~~~~
\nonumber \\
(\rho,~\theta)\longrightarrow(\rho-a_1 r \sin \rho /r~\sin 
\theta /r,~\theta+a_1 r\cos \rho /r~\cos \theta /r),\nonumber \\
(\rho,~\theta)\longrightarrow(\rho+a_2 r\sin \rho /r~\cos 
\theta /r,~\theta+a_2 r\cos \rho /r~\sin \theta /r),
\end{eqnarray}
where $(a_0,a_1, a_2) \in \dR^3 $ are small parameters. 

\noindent
The corresponding dynamical integrals (2.7) are
\begin{eqnarray}
J_0=p_{\theta}~r,~~~~~J_1=-p_\rho ~r\sin \rho /r~\sin 
\theta /r + p_\theta ~r\cos \rho /r~\cos \theta /r, \nonumber \\
J_2= p_\rho ~r\sin \rho /r~\cos \theta /r 
+ p_\theta ~r\cos \rho /r~\sin \theta /r,~~~~~~~~~~~~
\end{eqnarray}
where $p_\theta :=\partial L/\partial\dot{\theta},~p_\rho :=
\partial L/\partial\dot{\rho}$ are canonical momenta.

One can check that the dynamical integrals (2.12) satisfy the commutation 
relations of $sl(2,\dR)$ algebra
\begin{equation}
\{ J_0 , J_1 \} = -J_2,~~~\{ J_0 , J_2 \} = J_1,~~~\{ J_1 , J_2 \} = J_0.
\end{equation}

The constraint (2.6) reads  
\begin{equation}
(p_\rho^2 - p_\theta^2)\sin^2(\rho/r)=m^2.
\end{equation}
Making use of (2.12) we find that (2.14) relates the dynamical integrals  
\begin{equation}
J_1^2 + J_2^2-J_0^2  = \kappa^2,~~~~~~\kappa :=mr .
\end{equation}

\noindent
Eqs. (2.8) and (2.12) lead to equations for a particle trajectory
\begin{equation}
J_a y^a =0,~~~~~J_2 y^1 - J_1 y^2 =r^2 p_\rho ,
\end{equation}
where $p_\rho < 0,$ since we consider timelike trajectories. 

Each point $(J_0, J_1, J_2)$ of (2.15) defines uniquely a particle 
trajectory (2.16) on (2.4) admissible by the dynamics and consistent with 
the constraint (2.14). 
Thus, the one-sheet hyperboloid (2.15) defines the physical phase-space 
$\Gamma_h$. Since we intend to quantize the dynamics of a particle 
canonically, we should identify the symmetry group of $\Gamma_h$. At the 
level of space-time the symmetry group is $SO_0(1,2)$. The phase-space (2.15) 
and space-time  (2.4) have the same manifold structure, but it does not mean 
that $SO_0(1,2)$ is the only possible group action on $\Gamma_h$. Since the 
phase-space (2.15) is not simply connected, we can choose any group with 
$sl(2,\dR)$ as its Lie algebra to be the symmetry group, i.e. any covering 
group of $SO_0(1,2)$. This ambiguity is one of the sources of nonunigueness 
 of quantum dynamics considered in Sec. IIC.

\subsection{Observables}

We define classical observables to be smooth functions on 
 phase-space satisfying the following conditions: 
 (i) algebra of observables corresponds to the local symmetry of phase-space,  
(ii) observables specify particle trajectories admissible 
     by the dynamics ($V_p$ and $V_h$ are integrable systems), and 
(iii) observables are gauge invariant, i.e. have vanishing Poisson's 
             brackets with the constraint $G$, Eq. (2.6). 

In what follows we do not carry out the Hamiltonian reduction explicitly. 
We make use of our Hamiltonian reduction scheme to gauge invariant variables 
presented in \cite{GW2}. 

The canonical coordinates on phase-space are chosen in such a way that the 
classical observables are first order polynomials in one of the canonical 
coordinates. Such a choice enables, in the quantization procedure, solution 
of the operator-ordering problem by symmetrization. It also simplifies 
discussion of self-adjointness of quantum operators which in the linear case 
reduces to the solution of the first order linear differential equation 
(see, App. A). 

\subsection{Quantum dynamics}

In case the global symmetry of a classical system is defined by a Lie group 
with its Lie algebra being isomorphic to the Lie algebra of a local symmetry 
of the system, application of the GTQ method is straightforward. 
It consists in finding an irreducible unitary representation of the symmetry 
group on a Hilbert space. The representation space provides the quantum 
states space. The application of Stone's theorem \cite{MHS,MRS} to the 
representation of one-parameter subgroups of the symmetry group leads to 
self-adjoint operators representing quantum observables. 
Alternatively, by quantization we mean finding an essentially self-adjoint 
representation of the algebra of observables (corresponding to the local 
symmetry of the system) on a dense subspace of a Hilbert space, integrable 
to the irreducible unitary representation of the symmetry group of the 
gravitational system. 

Since our $V_h$ system satisfies the above symmetry relationship, it can be 
quantized by making use of our GTQ method. 

We choose $J_0,J_1$ and $J_2$ as the classical observables. 
One can easily verify that the criteria (i), (ii) and (iii) of Sec. IIB are 
satisfied. We parametrize the hyperboloid (2.15) as follows 
\begin{equation}
J_0=J,~~~J_1=J\cos\beta - \kappa\sin\beta,~~~J_2=J\sin\beta 
+\kappa\cos\beta ,
\end{equation}
where $~J\in \dR~$ and $~0\leq\beta <2\pi$.

\noindent
In this new parametrization the observables are linear in the coordinate $J$.

\noindent
One can check that the canonical commutation relation
$ \{J, \beta \}=1$ leads to Eq. (2.13).

Making use of the Schr\"{o}dinger representation for the canonical 
coordinates $J$ and $\beta$ (we set $\hbar =1$ through the paper) 

\[ \beta \rightarrow \hat{\beta}\psi(\beta):= \beta \psi(\beta) ,~~~
J \rightarrow \hat{J} \psi(\beta):= -i\frac{d}{d\beta}\psi(\beta), \]
and applying the symmetrization prescription to the products in (2.17) 

\[ J\cos\beta \rightarrow \frac{1}{2}(\hat{J}\cos\hat{\beta}+ 
\cos\hat{\beta}\hat{J}),~~~ J\sin\beta \rightarrow \frac{1}{2}(\hat{J}
\sin\hat{\beta}+ \sin\hat{\beta}\hat{J})\]
leads to 
\begin{equation}
\hat{J}_0\psi(\beta)= \Bigl[- i\frac{d}{d\beta}\Bigr]\psi(\beta) ,
\end{equation}
\begin{equation}
{\hat{J}}_1 \psi(\beta)= \Bigl[ \cos \beta 
~\hat{J_0} - ({\kappa} -\frac{i}{2})\sin \beta\Bigr]\psi(\beta),
\end{equation}
\begin{equation}
\hat{J}_2 \psi(\beta)= \Bigl[ \sin \beta 
~\hat{J_0} + ({\kappa} -\frac{i}{2})\cos \beta\Bigr]\psi(\beta) ,
\end{equation}
where $\psi \in\Omega_{\theta} \subset L^2(\dS)$,~ $ \theta \in \dR,~$ 
and where $L^2(\dS)$ is the space of square-integrable complex functions 
on a unit circle $\dS$ with the scalar product 
\begin{equation}
<\psi_1|\psi_2>:=\int_0^{2\pi} d\beta\overline{\psi_1(\beta)}\psi_2(\beta).
\end{equation}

The subspace $ \Omega_{\theta}$ is defined to be 
\begin{equation}
\Omega_{\theta}:=\{\psi\in L^2(\dS)~|~\psi\in C^\infty [0,2\pi],~\psi^{(n)}
(0)=e^{i\theta}\psi^{(n)}(2\pi),~n=0,1,2,...\}.
\end{equation}
The representation  (2.18 - 2.22) is parametrized by $\theta \in \dR$.

The unbounded operators $\hat{J}_a~~(a=0,1,2)~$ are well defined because 
$ \Omega_{\theta}$ is a dense subspace of the Hilbert space $L^2(\dS)$.

It is clear that $\Omega_{\theta}$ is a common invariant  domain for 
all $ \hat{J}_a$ and their products.

\noindent
One can verify that 
\begin{equation}
[\hat{J}_a,\hat{J}_b]\psi =-i\widehat{\{J_a,J_b\}}\psi,
~~~~\psi\in\Omega_{\theta} ,
\end{equation}
and that the representation (2.18 - 2.22) is symmetric on $\Omega_{\theta}$. 
We prove in  App. A that this representation is essentially self-adjoint.

It is known \cite{GW2,Loll,Ply} that the parameter $\theta$ labels unitarily 
non-equivalent representations of $sl(2,\dR)$ algebra corresponding to the 
unitary representations of various covering groups of $SO_0(1,2)$, so it is 
connected with the ambiguity in the choice of the symmetry group of phase-space 
considered at the end of Sec. IIA. For the purpose of clarity of discussion 
of the new ambiguity problem connected directly with singularities of 
space-time of $V_p$ system, we carry out further discussion for a fixed 
value of $\theta$. In what follows we put $\theta =0$ which corresponds to 
the choice of $SO_0(1,2)$. The ambiguities in quantization connected with 
the choice of $\theta$ and with the choice of a symmetry group of phase-space 
in general will be considered elsewhere \cite{WJP}.

The problem of finding representations of the group $SO_0(1,2)$ was  
considered in 1947 by Bargmann \cite{VVB} in the context of representation 
of $SU(1,1)$ group. There exists two-to-one homomorphism of $SU(1,1)$ group 
onto $SO_0(1,2)$ with the kernel  $\dZ_2 :=\{e, -e\}$, where $e$ is the 
identity element of $SU(1,1)$. Thus, the factor group $SU(1,1)/\dZ_2$ is 
isomorphic to $SO_0(1,2)$.

Bargmann has constructed and classified all irreducible unitary 
representations of $SU(1,1)$ group by making use of the multiplier 
representation method \cite{VVB,NVK}. These representations fall basically 
into three classes \cite{VVB,PJS}: principal series, complementary series and 
discrete series. Bargmann's classification is based on: (i) his special 
decomposition of $SU(1,1)$ group (see, Eq. (4.12) of \cite{VVB}) 
into a product 
of one-parameter subgroups one of which is a compact Abelian group with 
unitary representation having complete system of vectors and integral 
(corresponding to $SO_0(1,2)$ group) or half-integral proper values, and 
(ii) his classification of irreducible representations of $su(1,1)$ algebra. 

To compare our representation within Bargmann's, we  choose  
as a basis in $L^2(\dS)$ his basis
\begin{equation}
\phi_m(\beta)=(2\pi)^{-1/2}\exp(im\beta),~~~0\leq \beta <2\pi,~~~m \in 
\dZ:=\{0,\pm1,\pm2,...,\}.
\end{equation}

Since the algebras $so(1,2),~su(1,1)$ and $sl(2,\dR)$ are isomorphic 
\cite{NVK}, we make the comparison with Bargmann's representation at the level 
of algebra. Correspondingly, we examine the action of the operators 
$\hat{J}_a$ and $\hat{C}$ on the subspace $\Omega :=\Omega_{\theta =0}$ 
spanned by the set of 
vectors (2.24). The operator $\hat{C}$ corresponds to the Casimir operator 
$C$ of $sl(2,\dR)$ algebra. $C$ is defined to be \cite{MKW}
\begin{equation}
C= J_1^2 +J_2^2 - J_0^2.
\end{equation}
In our representation the operator $\hat{C}$ reads
\begin{equation}
\hat{C}\psi= [\hat{J}_1^2 +\hat{J}_2^2 - \hat{J}_0^2]\psi =
(\kappa^2 +1/4)\psi ,~~~~\psi \in \Omega ,
\end{equation}
where the third term in (2.26) was obtained by making  use of explicit 
formulas for $\hat{J}_a ,$ Egs. (2.18 - 2.20).

It is easy to verify that the action of the operators 
$\hat{J}_a$ on $\phi_m $ reads
\begin{equation}
\hat{J}_0 \phi_m =m \phi_m,~~~~~m\in \dZ 
\end{equation}
\begin{equation}
\hat{J}_1 \phi_m =\frac{1}{2} (m+1/2+i\kappa)\phi_{m+1}+
\frac{1}{2} (m-1/2-i\kappa)\phi_{m-1},
\end{equation}
\begin{equation}
\hat{J}_2 \phi_m =-\frac{i}{2} (m+1/2+i\kappa)\phi_{m+1}+
\frac{i}{2} (m-1/2-i\kappa)\phi_{m-1} .
\end{equation}

At this stage we are ready to discuss the connection of our representation 
with Bargmann's.
Direct comparison of Eqs. (2.26 - 2.29) with Bargmann's (6.14), (6.21) and 
(6.22) of Ref. \cite{VVB} shows that the following identification is possible: 
\begin{equation}
\hat{C} \equiv Q,~~\kappa^2 \equiv q-1/4,~~\hat{J}_0 \equiv H_0=i\Lambda_0,
~~\hat{J}_1 \equiv -H_1 =-i\Lambda_1,~~\hat{J}_2 \equiv -H_2=-i\Lambda_2 ,
\end{equation} 
where $Q,q,H_a,\Lambda_a~~(a=0,1,2)$ are Bargman's quantities used to define 
his representation of $su(1,1)$ algebra.

The range of our parameter $~\kappa =mr~$ is $~0<\kappa <\infty, ~$ so it 
corresponds to Bargmann's $~1/4 <q < \infty.~$ Therefore, our 
representation is almost everywhere identical with  Bargmann's continuous 
class integral case (corresponding to $SO_0(1,2)$ group) called $~C_q^0~$ 
with $~1/4 \leq q < \infty$, which is also called  the principal series of 
irreducible unitary representation of $SU(1,1)$ group \cite{PJS}. 
The only difference is that for massive particle $~m>0~$, thus 
$\kappa = mr > 0 $, so $~q > 1/4 .~$ 
The precise identity may occur, if taking the limit $~\kappa \rightarrow 0~$ 
can be given physical and mathematical sense in our formalism. We discuss 
this issue  in Sec. IVA.

\section{Particle on plane}

\subsection{Restrictions for classical dynamics}

The Lagrangian (2.5) with the metric tensor defined by (2.2) reads
\begin{equation}
L=-m\sqrt{\dot{t}^2- \dot{x}^2\exp(2t/r)}, 
\end{equation}
where $t:=x^0,~x:=x^1,~\dot{t}=dt/d\tau$ and $\dot{x}=dx/d\tau$.

\noindent
The local symmetries of $L$ (and the infinitesimal transformations of 
$V_p$ space-time) are defined by translations
\begin{equation}
(t,~x)\longrightarrow (t,~x+b_0),
\end{equation}
space dilatations with time translations
\begin{equation}
(t,~x)\longrightarrow (t-rb_1,~x+xb_1 ) ,
\end{equation}
and by the transformations
\begin{equation}
(t,~x)\longrightarrow (t-2rxb_2 ,~x+ (x^2 +r^2 e^{-2t/r})b_2),
\end{equation}
where $(b_0,b_1,b_2) \in \dR^3 $ are small parameters.

\noindent
The Killing vector fields corresponding to the transformations (3.2 - 3.4) 
define, respectively, the dynamical integrals (2.7)
\begin{equation}
 P=p_x,~~~K=-rp_t+xp_x,~~~M=-2rxp_t +(x^2 + r^2 e^{-2t/r}) p_x ,
\end{equation}
where $p_x = \partial L/\partial\dot{x},~ p_t = \partial L/\partial \dot{t}$. 

\noindent
One can verify that the dynamical integrals (3.5) satisfy the commutation 
relations of $sl(2,\dR)$ algebra in the form
\begin{equation}
\{P,K\}=P,~~~\{K,M\}=M,~~~\{P,M\}=2K.
\end{equation}

\noindent
The mass-shell condition (2.6) takes the form
\begin{equation}
p^2_t -e^{-2t/r}p^2_x =m^2,
\end{equation}
 which, due to (3.5), relates the dynamical integrals   
\begin{equation}
K^2-PM=\kappa ^2,~~~~~ \kappa =mr.
\end{equation}

By analogy to $V_h$ case one may expect that each triple $(P,K,M)$ 
satisfying (3.8) determines a trajectory of a particle.  However, not all 
such trajectories are consistent with particle dynamics:

\noindent
For $P=0$ there are two lines $ K=\pm \kappa$ on the hyperboloid (3.8).
Since by assumption $\dot{t} >0$, we have  $p_t = \partial L/
\partial\dot{t}=-m\dot{t}\:(\dot{t}-\dot{x}\exp(2t/r))^{-1/2} < 0 $. 
According to (3.5) $K-xP = -rp_t,$ thus $K-xP > 0,~$ i.e. $K>0~$ for $P=0$.
Therefore, the line $(P=0,~K=-\kappa)$ is not available for the dynamics. 
The hyperboloid (3.8) without this line defines the physical phase-space 
$\Gamma_p$, which is topologically equivalent to $\dR^2$.
 
Excluding the momenta $p_t$ and $p_x$ from (3.5) we find explicit formulae 
for particle trajectories 
\begin{equation}
x(t)=M/2K,~~~~~\mbox{for}~~~~P=0
\end{equation}
and 
\begin{equation}
x(t)=\Bigl[K-\sqrt{\kappa^2 +(rP)^2\exp{(-2t/r)}}\Bigr]/P,
~~~~~\mbox{for}~~~~P\neq 0,  
\end{equation}
where (3.10) takes into account that $K-x P>0.$

The space of trajectories defined by (3.9) and (3.10) represents the  
phase-space $\Gamma_p$.

\subsection{Choice of observables}

To satisfy all required criteria for observables, we parametrize 
$\Gamma_p$ by the coordinates $(q,p)\in \dR^2 $  as follows
\begin{equation}
P=p,~~~K=pq-\kappa,~~~M=pq^2 -2\kappa q.
\end{equation}
The integrals (3.11) satisfy the algebra (3.6), if $~\{p,q\}=1$.

To compare quantum dynamics of $V_p$ and $V_h$ systems, let us bring 
their observables  to the same functional form. 
 It can be achieved in two steps: 

\noindent
First, we change parametrization of the phase-space $\Gamma _p$ as follows
\begin{equation}
q=:\cot\frac{\sigma}{2},~~~p=:(1-\cos\sigma)(I+\kappa\cot\frac{\sigma}{2}),
\end{equation}
where $0<\sigma<2\pi$ and $I\in \dR$.

\noindent
Second, we rewrite the observables (3.11) in terms of new canonical 
variables $(I,\sigma)$ and redefine them. The final result is 
\begin{equation}
I_0:=\frac{1}{2}(M+P)=I,
\end{equation}
\begin{equation}
I_1:=\frac{1}{2}(M-P)=I\cos\sigma -\kappa\sin\sigma,
\end{equation}
\begin{equation}
I_2:=K=I\sin\sigma +\kappa\cos\sigma.
\end{equation}

Since $\{\sigma, I\}=1$, the commutation relations for $I_a$ resulting from 
(3.6) are identical to the commutation relations (2.13) for $J_a ~ (a=0,1,2)$.

Comparing (3.13 - 3.15) with (2.17) we can see that $I_a$ and 
$J_a~~(a=0,1,2)$ have the same functional forms, but they are different 
because the range of parameter $\beta$ is $0\leq \beta < 2\pi$, whereas 
the range of $\sigma$ reads $0< \sigma < 2\pi $. This difference results 
from the difference between the topologies of phase-spaces of $V_h$ and $V_p$ 
systems: $\Gamma_h$ is the hyperboloid (2.15), whereas $\Gamma_p$ is the 
hyperboloid (3.8) without one line.
Therefore, the phase-space $\Gamma_p$ cannot be invariant under the action 
of $SO_0(1,2)$ group. This may be already seen in the context of space-times, 
since $V_p$ is only a subspace of $V_h$ due to the isometric 
immersion map (2.3).  In fact, the Killing vector field generated by the 
transformation (3.4) is not complete on $V_p$, whereas the vector fields 
generated by (3.2) and  (3.3) are well defined globally (see, App. B).   
Therefore, the dynamical integral $M$ is not well defined globally. 
Let us make the assumption that each classical observable should be 
a globally well defined function on a physical phase-space. Then, the set of 
observables of $V_p$ system consists of only the integrals $P$ and $K$ 
satisfying the algebra (see, (3.6))
\begin{equation}
\{P,K\} = P.
\end{equation}
Eq. (3.16) defines a solvable subalgebra of $sl(2,\dR)$ algebra. 

The algebra (3.16) is isomorphic  to the  algebra $~\aff(1,\dR)~$ of the 
affine group $~\Aff(1,\dR)$. 
If we denote the span of the algebra (3.16) by $<P,K>$ and the span of 
$\aff(1,\dR)$ by $<A,B>$, the algebra isomorphism is defined by $A:=-K$ 
and $B:=-P$. The algebra $\aff(1,\dR)$ is defined by the commutation relation 
\begin{equation}
\{A,B\} = B.
\end{equation}
One can easily show that the center of $\Aff(1,\dR)$ is an identity element 
of this group. Thus, $\Aff(1,\dR)$ is the only Lie group with $\aff(1,\dR)$ 
as its Lie algebra. This circumstance makes unique the choice of the 
symmetry group of the phase-space $\Gamma_p$.

The algebra $~\aff(1,\dR)~$ is quite different from  
the local symmetry (3.6) of $\Gamma_p$. 
The  relationship between local and global symmetries which occurs in $V_h$ 
case does not exist in the present case.

\subsection{Quantum dynamics on plane}

In  gravitational systems the global and local symmetries may easily happen 
to be incompatible. An example is our $V_p $ system of a free particle in 
space-time with removable type singularities.  In such cases our GTQ 
method needs modification to be applicable. We propose to complete the set of 
conditions defining the algebra of observables, Sec. IIB, by the following 
one:  \textit{(iv) algebra of observables is consistent with the global 
symmetry of  phase-space}.
 
To quantize the algebra (3.17) we use tha fact that all unitary irreducible 
representations of  $\Aff(1,\dR)$ group are known. They were discovered 
already in 1947  by Gel'fand and Najmark \cite{GFN}.
In what follows we adopt the Vilenkin version  \cite{NVK}. 
There exist only two (nontrivial) unitarily non-equivalent representations
\begin{equation}
U_s(g):\mathcal{H}_\Lambda \longrightarrow \mathcal{H}_\Lambda ,~~~~~s=-,+
\end{equation}
where $g\in \Aff(1,\dR)$ and where $\mathcal{H}_\Lambda$ is the Hilbert space 
defined as follows:

\noindent
We introduce the space $\Lambda$
\begin{equation}
\Lambda:= C^\infty_0(\dR_+),~~~~\dR_+:=\{x\in \dR~|~x>0\},
\end{equation}
with the scalar product given by
\begin{equation}
<\varphi_1| \varphi_2> :=\int^\infty_0 \overline{\varphi_1(x)}\varphi_2 
(x) \frac{dx}{x},~~~~~~\varphi_1 , \varphi_2 \in \Lambda .
\end{equation} 
$\mathcal{H}_\Lambda$  is the Hilbert space obtained by completion of 
$\Lambda$  with respect to  the scalar product (3.20). The operators 
$U_s(g)$ are defined as follows \cite{NVK}
\begin{equation}
U_s[g(a,b)]\psi(x):= \exp(-isbx)~\psi(ax),~~~~\psi \in \mathcal{H}_\Lambda ,
~~~~s=-,+
\end{equation}
where $g(a,b)\in \Aff(1,\dR)$ and $(a,b)\in \dR_+ \times \dR,~$  
 parametrize the  group elements.

\noindent
It is easy to check that  (3.21) is a representation of $\Aff(1,\dR)$ group. 

Since the measure $x^{-1}dx$ in (3.21) is invariant with respect to 
$x\rightarrow ax$, we obtain
\begin{equation}
<U_s[g(a,b)]\psi_1~|~U_s[g(a,b)]\psi_2>=\int^\infty_0\overline
{\psi_1(ax)}\psi_2(ax)\frac{dx}{x}=
\int^\infty_0 \overline{\psi_1(x)}\psi_2(x)\frac{dx}{x}=<\psi_1|\psi_2>
\end{equation}
for all $\psi_1 , \psi_2 \in \mathcal{H}_\Lambda$, which shows that (3.21) 
defines a unitary representation. Making use of the reasoning of Ref. 
\cite{NVK} one can prove that the unitary representation (3.21) is 
irreducible.

The application of Stone's theorem (strong form) to (3.21) defines two sets 
of operators $\hat{A}_s$ and  $\hat{B}_s~~(s=-,+).$
\begin{equation}
\frac{d}{dt}_{\mid_{t=0}}U_s[g(a(t),0)]\varphi (x)=x\frac{da(0)}{dt}
\frac{d\varphi(x)}{dx} = x\frac{d}{dx}\varphi(x) = i(-i x\frac{d}{dx}
\varphi(x)) =: i\hat{A}_s\varphi(x)
\end{equation}
and 
\begin{equation}
\frac{d}{dt}_{\mid_{t=0}}U_s[g(1,b(t))]\varphi (x)=
-isx\frac{db(0)}{dt}e^{-isb(0)x}\varphi (x)= 
i(-sx)\varphi(x)=: i\hat{B}_s\varphi(x),
\end{equation}
where $~t \rightarrow a(t)~$ and $~t \rightarrow b(t)~ $  with the boundary 
conditions $~a(0)=1,~~da(0)/dt=1$ and $~b(0)=0,~~db(0)/dt=1,~$ respectively, 
are two integral curves on $~\Aff(1,\dR).$

Equations (3.23) and (3.24) define the domains for the operators $\hat{A}_s$ 
and $\hat{B}_s$. One can prove (see, App. C) that these operators are 
essentially self-adjoint on the space $\Lambda$ defined by (3.19). 
One can also verify that
\begin{equation}
[\hat{A}_s,\hat{B}_s]\varphi=-i\hat{B}_s\varphi,~~~~\varphi \in \Lambda ,
~~~~s=-,+
\end{equation}
which demonstrates that (3.25) is the representation of the algebra 
$\aff(1,\dR)$ defined by (3.17).

Therefore, there are possible only two (up to unitary equivalence)  
quantum dynamics corresponding to a single classical dynamics of $V_p$ 
system. We can use either the representation $U_+(g)$ or $U_-(g)$. 
(We ignore the trivial one-dimensional representation mentioned in 
\cite{NVK}.) 

The quantization of $V_p$ system is now complete.

To appreciate the quantization requirement that representation of the 
algebra of observables should be integrable to the unitary representation 
of the symmetry group of the system, let us consider again 
the representation of 
$sl(2,\dR)$ algebra satisfied by $I_a ~(a=0,1,2)~$ observables (3.13 - 3.15). 
Since $I_a$ and $J_a ~(a=0,1,2)~$ have the same functional forms and have 
almost everywhere the same ranges, the representation of $I_a$ observables 
is defined by (2.18 - 2.22) with $\hat{J}_a$ replaced by $\hat{I}_a$ and 
$\beta$ replaced by $\sigma$. 
However, there exists no value of the parameter $\theta$ in (2.22) which can 
lead to the algebra representation integrable to the unitary representation 
of the symmetry group $\Aff(1,\dR)$. It is so because $sl(2,\dR)$ is not 
the algebra of $Aff(1,\dR)$.

\section{Dynamics of  massless particle} 

\subsection{Massless particle on hyperboloid}

To obtain the description of dynamics of a massless particle on 
hyperboloid we examine taking the limit $\kappa \rightarrow 0$, i.e.
$m\rightarrow 0$, in Sec. II. 
The inspection of classical and quantum dynamics 
of $V_h$ system reveals that apart from Eq. (2.5) for the Lagrangian, all 
equations can be considered in the limit  $m\rightarrow 0$: 

\noindent
The phase-space $\Gamma_h$ defined by (2.15) turns into two cones 
\begin{equation}
J_1^2 +J_2^2 -J_0^2 =0.
\end{equation}
with a common vertex $V$ defined by $J_0 =0=J_1 =J_2$.

\noindent
It is clear that each cone is invariant under the action of $SO_0(1,2)$ 
group. 

Each point of (4.1) different from $V$ labels  uniquelly the  trajectory  
of a particle on hyperboloid (2.4). The set of trajectories (stright lines) 
is the set of generatrices of the hyperboloid (2.4).

Parametrizing (4.1) by $~J_a~$ in the form (2.17) with $\kappa =0$ leads to 
(2.18 - 2.20) with $\kappa =0$ as well. 

The quantum Casimir operator (2.26) now reads 
\begin{equation}
\hat{C}\psi =\frac{1}{4}\psi,~~~~\psi\in \Omega.
\end{equation} 

There is no problem with taking  $\kappa \rightarrow 0$  in (2.27 - 2.29) too. 
The only problem is the form of the Lagrangian (2.5) because $m$ occurs as 
a factor. 
We can avoid this difficulty by choosing the  Lagrangian which does not 
depend explicitly on the mass of a particle \cite{BDV,GW4}
\begin{equation}
\mathcal{A}=\int_{\tau_1}^{\tau_2}~L(\tau)~d\tau,~~~~~L(\tau)
:=-\frac{1}{2\lambda(\tau)}g_{\mu\nu}(x^0(\tau),x^1(\tau))\;\dot{x}^\mu (\tau)
\dot{x}^\nu (\tau),
\end{equation}
where $~\tau~$ is an evolution parameter, $\dot{x}^\mu = dx^\mu/d\tau~$ 
and $~\lambda~$ plays the role of Lagrangian multiplier. The action (4.3) 
is invariant under reparametrization $\tau\rightarrow f(\tau),~~\lambda(\tau)
\rightarrow\lambda(\tau)/\dot{f}(\tau).$ This gauge symmetry leads to dynamics 
constrained by (2.14) with $m=0$ and consequently to (2.15) with $\kappa =0,$ 
i.e. to Eq. (4.1). Thus the dynamics of a massless particle defined by (2.5) 
and (4.3) are equivalent. It appears that the massless particle dynamics 
of quantum $V_p$ system may be described by the continuous (integral case) 
Bargmann's $~C^0_q~$ class with $q=1/4$ (see, the last paragraph of Sec. IIC). 

Thus, the principal series irreducible unitary representation of 
$SO_0(1,2)$ group appears to be able to describe quantum dynamics of 
both massive and massless particle on hyperboloid.

However, we have not analysed the vertex of (4.1) in the context of  particle 
dynamics carefully enough. The subtlety is that in case of a massless particle 
the vertex cannot be used to specify the trajectory of a particle 
by making use of (2.16). For $p_\rho \neq 0$ the Eq. (2.16) has no solutions 
and for $p_\rho = 0$ it has infinitely many.
The removal of the vertex from (4.1) turns the phase-space of 
$V_p$ system into two separate cones $\mathcal{C}_+$ and $\mathcal{C}_-$ 
with $J_0 >0$ and $J_0 <0$, respectively. This procedure splits the system 
into two parts and each part can be quantized independently. 
The corresponding quantum systems have been already found \cite{GW4}: the 
quantum system connected with $\mathcal{C}_+$ may be described by the 
discrete series $D_+$ (having positive spectrum of $\hat{J}_0$) of the 
irreducible unitary representation of $SO_0(1,2)$ group, whereas the 
discrete series $D_-$ (with negative spectrum of $\hat{J}_0$) may be used to 
represent the quantum system connected with $\mathcal{C}_-$. Thus the whole 
quantum system may be described by $D_- \oplus D_+ $ representation 
(see, Sec. 2 of \cite{GW4} for more details).

The role of the vertex $V$ is of primary importance. If we take it to belong 
to the phase-space, the corresponding quantum system will be unique and may 
be described by an irreducible representation of $SO_0(1,2)$ group. 
The phase-space without the vertex leads to infinitely many reducible 
representations of $SO_0(1,2)$. Therefore, taking $\kappa\rightarrow 0$  
at the classical level leads to the latter. Taking the limit not at the 
classical, but at the quantum level gives the former.

\subsection{Massless particle on plane}

In case of dynamics on plane (see, Sec. III) an action integral is defined 
by (4.3) and the one-sheet hyperboloid (3.8), in the limit $m\rightarrow 0,$ 
turns into `one-sheet cone' 
\begin{equation}
K^2-PM=0.
\end{equation}
Since the dynamics requires $K>0$ for $P=0$, we have to remove the line 
$(P=0=K)$ from (4.4) to get the physical phase-space $\Gamma_p$. 
As the result, the phase-space slits into two disconnected parts 
$\mathcal{P}_+$ and $\mathcal{P}_-$ with $P>0$ and $P<0$, respectively.
The observables $P$ and $K$ are well defined globally on  $\mathcal{P}_+$ 
and $\mathcal{P}_-$, and the corresponding $A$ and $B$ observables satisfy 
 $\aff(1,\dR)$ algebra (3.17) with $\Aff(1,\dR)$ as the symmetry group. 

Since at the quantum level there is no explicit dependence on the parameter 
$\kappa$, taking $m\rightarrow 0$ is trivial. 

Therefore, the quantum dynamics on plane of the massless  particle 
may be described  by the representation $T_-\oplus T_+$, where $T_-$ and 
$T_+$ denote the representations corresponding to $s=-$ and $s=+$, 
respectively. 

In the present case there is no ambiguity connected with the vertex of (4.4) 
since the removed line (P=0=K) includes the vertex.

\section{Conclusions}

Test particle is a tool which may be used in the theory of classical gravity 
to examine causal geodesics of singular space-times. We have tried to quantize 
the particle dynamics to see what can one do to avoid problems connected with 
\textit{removable} singularities of space-time in quantum theory. 
Our main result is that taking account of \textit{global} 
properties of space-time makes possible the imposition of quantum rules into 
the dynamics of a particle. It is clear that we have not tried to quantize 
gravitational field, but only the dynamics of a particle.

Local properties of a given space-time described by metric tensor and Lie 
algebra of the Killing vector fields do not specify the system uniquely 
because space-times with  different global properties may have isometric Lie 
algebras \cite{LML,LPP}. Presented results show that the topology of 
space-time carries the 
information not only on the symmetry group but also indicates which local 
properties of the  system should be used in the quantization procedure. 
Our results are  consistent with the fact that quantum theory is a global 
theory in its nature. We suggest that its consolidation with classical gravity 
should take into account both local and global properties of space-time.
The Einstein equations being partial differential equations cannot specify 
the space-time topology, but only its local properties. Fortunately, the 
mathematics of low dimensional manifolds offers a full variety of topologies 
for space-time models consistent with local properties of a given space-time 
\cite{WPW,WPT}.

Generalization of our results to the four-dimensional 
de Sitter space-times seems to be  straightforward.
The space-time with topology  $\dR^1 \times \dR^3 $, the four dimensional 
analog of $V_p$, is geodesically incomplete and it can be embedded 
isometrically \cite{SHE} into the space-time with topology  
$\dR^1 \times \dS^3$, corresponding to $V_h$, by generalization of the 
mapping (2.3). The quantum dynamics of a particle on four dimensional 
hyperboloid in five dimensional Minkowski space is presented in \cite{WG9}. 
It seems that quantization of dynamics of a particle on de Sitter space-time with topology 
$\dR^1 \times \dR^3 $ may be carried out by analogy to the quantization 
of $V_p$ system.  
First of all one should find the set of all Killing vector fields which are 
complete. They would help to identify the algebra of  globally well defined
dynamical integrals and the symmetry group of phase-space. 
Unitary representations of the symmetry group may be used to define quantum 
dynamics of a particle. It is clear that examination of dynamics of a particle 
in four dimensional space-time is much more complicated than in two dimensional 
case, but it should be feasible if the number of globally well defined 
dynamical integrals is high enough. We expect that application of our method 
to the four dimensional case should lead to the conclusion similar in its 
essence to the two dimansional case.

Our paper concerns removable type singularities of space-time. 
Great challenge is an extension of our analysis to space-times with 
essential type singularities, i.e. including not only  incomplete geodesics,  
but also blowing up Riemann tensor components or  curvature 
invariants \cite{JMS}. The FLRW type universes appear to be good candidates to 
begin with, since their  properties are well known \cite{GFR}. 
Our method of analysing particle dynamics by making use of embeddings into 
the Minkowski space extends to higher dimensions. There exist theorems of 
differential geometry \cite{LPE,JJP,PSW,AB1}  that every curved 
four-dimensional space-time can be embedded isometrically into a flat 
pseudo-Euclidean space $E_N$ with $5\leq N\leq 10$.

Recently, Heller and Sasin put forward the idea of modeling space-time 
by the  Connes noncommutative geometry. With this new idea one can try 
to coup with  space-time singularities and try to establish the relationship 
with quantum description \cite{MH1,MH2}.

Completely different approach has been developed by  Ashtekar and his 
collaborators (see, \cite{AAA} and references therein). This non-perturbative 
and background-independent theory of quantum gravity seems to be 
free of problems connected with space-time singularities \cite{MB1,MB2}, 
and it seems to reproduce (in its classical limit) the Einstein theory 
of gravity. 

Finally, let us make some general comments concerning the ambiguity problems 
of the canonical quantization procedure. It is known that quantization of 
a system with constraints is a highly nonunique procedure. One may quantize 
first and then impose the constraints, or vice versa. Having fixed the 
phase-space one may quantize geometrically, group theoretically, by making use 
of coherent states or by some mixture of all these methods. 
Only quantization of the simplest mechanical systems leads to similar or 
identical results. The ambiguity may be reduced by making use of global 
properties of classical system. In principle, the ambiguity should be 
removed by comparison of predictions with experimental data. 
In case of the systems with a few degrees of freedom and with non-trivial 
topology of phase-space the  discussion of the ambiguity problem 
was recently done in Refs. \cite{MB3,MB4}. An extension of this 
discussion will be published elsewhere \cite{WJP}.

\appendix

\section{Representation algebra on hyperboloid}

 Let $L^2(\dS)$ denotes the Hilbert space of square integrable complex 
functions on a unit circle with the inner product
\begin{equation}
<\varphi|\psi>=\int_{0}^{2\pi}d\beta~\overline{\varphi(\beta)}\psi(\beta),
~~~~~~~~\varphi,\psi\in L^2(\dS).
\end{equation} 

In what follows we outline the prove that  representation of $sl(2,R)$ 
algebra defined by
\begin{equation}
\hat{J}_0\psi(\beta):=\frac{1}{i}\frac{d}{d\beta}\psi(\beta),~~~\beta 
\in \dS,~~~\psi\in\Omega_{\theta}~~~\theta\in \dR ,
\end{equation}
\begin{equation}
{\hat{J}}_1 \psi(\beta):= \Bigl[ \cos \beta 
~\hat{J_0} - ({\kappa} -\frac{i}{2})\sin \beta\Bigr]\psi(\beta),
\end{equation}
\begin{equation}
\hat{J}_2 \psi(\beta):= \Bigl[ \sin \beta 
~\hat{J_0} + ({\kappa} -\frac{i}{2})\cos \beta\Bigr]\psi(\beta) ,
\end{equation}
where 
\begin{equation}
\Omega_{\theta} := \{\psi\in L^2(\dS)~|~\psi\in C^\infty[0,2\pi],
~\psi^{(n)}(0)= e^{i\theta}\psi^{(n)}(2\pi),~ n=0,1,2...\},
\end{equation}
is essentially self-adjoint.

It is clear that $\Omega_{\theta}$ is a dense invariant common domain for 
$\hat{J}_a~~(a=0,1,2)$. Since the functional form of $\hat{J}_a$ does not 
depend on $\theta$ and since $\exp(-i\theta)\cdot\exp(i\theta)=1$, the 
operators are symmetric on $\Omega_{\theta}$: 

\noindent
An elementary proof includes integration 
by parts of one side of 
\begin{equation}
<\phi_1|\hat{J}_a\phi_2>=<\hat{J}_a \phi_1|\phi_2>,~~~~\phi_1,\phi_2 
\in \Omega_{\theta}
\end{equation}
followed by making use of the property
\begin{equation}
\phi(0)=\exp(i\theta) \phi(2\pi),~~~~\phi\in \Omega_{\theta} .
\end{equation}

The domains $D(\hat{J}_a^\ast)$ of the adjoint $\hat{J}_a^\ast$ of 
$\hat{J}_a$ consists of functions $\psi_a$ which  satisfy the condition 

\begin{equation}
\psi_a(0)=\exp(i\theta)\psi_a(2\pi),~~~~\psi_a \in D(\hat{J}_a^\ast)
\subset L^2(\dS)
\end{equation}
for $a=0,1,2$. 

The main idea of the proof \cite{MRS} is to show that the only solutions 
to the equations
\begin{equation}
\hat{J^\ast_a}f_{a\pm} = \pm if_{a\pm},~~~~f_{a\pm}\in D(J^\ast_a),~~~~
a=0,1,2
\end{equation}
are $f_{a\pm}(\beta)=0$, i.e. the deficiency indices of $\hat{J_a}$ on 
$\Omega_{\theta}$  satisfy $n_{a+}=0=n_{a-}~$ (for $a=0,1,2 $).

The equation (A9) for $a=0$ reads
\begin{equation}
\frac{1}{i}\frac{d}{d\beta}f_{0\pm}(\beta)=\pm if_{0\pm}(\beta)
\end{equation} 
and its general normalized solution is
\begin{equation}
f_{0\pm}(\beta)=C_{0\pm} \exp(\mp\beta),~~~~C_{0+}:=\sqrt{2/(1-\exp(-4\pi))},
~~~~C_{0-}:=\sqrt{2/(\exp(4\pi)-1)}.
\end{equation}

\noindent
The solutions (A11) does not satisfy (A8). Thus the only solution to (A10) 
is $f_{0\pm}(\beta)=0$.

For $a=1$ the equation (A9) can be written as
\begin{equation}
(\cos \beta\frac{d}{d\beta}-r\sin\beta+\lambda_\pm)f_{1\pm}(\beta)=0 ,
\end{equation}
where $r=1/2+\kappa i,~ \kappa\in \dR,~ \lambda_\pm =1$ or $-1$ for $f_{1+}$ 
or $f_{1-}$, respectively.

\noindent
One can verify that the general solution of (A12) reads
\begin{equation}
f_{1\pm}(\beta)= C_{1\pm}|\cos\beta|^{-r}|\tan(\frac{\beta}{2}+
\frac{\pi}{4})|^{-\lambda\pm} ,
\end{equation}
where $C_{1\pm}$ are complex constants.

\noindent
The immediate calculations show that for $C_{1\pm}\neq 0$
\begin{equation}
\lim \Re f_{1+}(\beta)=\infty =\lim \Im f_{1+}(\beta)~~~\text{as}~~~
\beta\rightarrow \frac{3}{2}\pi\pm
\end{equation}
and
\begin{equation}
\lim \Re f_{1-}(\beta)=\infty =\lim \Im f_{1-}(\beta)~~~\text{as}~~~
\beta\rightarrow \frac{\pi}{2}\pm .
\end{equation}
Therefore $f_{1\pm}$ are not square integrable  and the only 
solutions of (A12) are $f_{1\pm}=0$.

The equation (A9) for $a=2$ has the form
\begin{equation}
(\sin\beta\frac{d}{d\beta}+r\cos\beta +\lambda_\pm)f_{2\pm}(\beta)=0 ,
\end{equation}
where $r=1/2 +\kappa i$ and $\lambda\pm =1$ or $-1$, for $f_{2+}$ or 
$f_{2-}$, respectively. 

\noindent
The general solution to (A16) is
\begin{equation}
f_{2\pm}(\beta)=C_{2\pm}|\sin\beta|^{-r}|\tan\frac{\beta}{2}|^
{-\lambda\pm} ,
\end{equation}
where $C_{2\pm}$ are complex constants.

\noindent
The standard calculations yield 
\begin{equation}
\lim \Re f_{2+}(\beta)=\infty =\lim \Im f_{2+}(\beta)~~~\text{as}~~~
\beta\rightarrow 0+~~~\text{or}~~~\beta\rightarrow 2\pi -  
\end{equation}
and
\begin{equation}
\lim \Re f_{2-}(\beta)=\infty =\lim \Im f_{2-}(\beta)~~~\text{as}~~~
\beta\rightarrow \pi\pm .
\end{equation}
Thus,  $f_{2\pm}$ are not square integrable unless $C_{2\pm} =0.$ 

This finishes the proof, the detailed verification of consecutive 
steps being left to the reader.

\section{Global transformations on plane}

The transformations (3.2), (3.3) and (3.4) of Sec. IIIA lead, respectively, 
to the following infinitesimal generators  
\begin{equation}
X_1 = \partial/\partial x,
\end{equation}
\begin{equation}
X_2 = -r\partial/\partial t + x \partial/\partial x,
\end{equation}
\begin{equation}
X_3 = -2rx\partial/\partial t + (x^2 + r^2 \exp(-2t/r))\partial/\partial x .
\end{equation}
The one-parameter group generated by $X_3$ is defined by the solution 
of the Lie equations 
\begin{equation}
\frac{dt}{db_3}=-2rx,
\end{equation}
\begin{equation}
\frac{dx}{db_3}=x^2 +r^2 \exp(-2t/r) ,
\end{equation}
\begin{equation}
t_{\mid b_1 =0= b_2 =b_3} =t_0
\end{equation}
\begin{equation}
x_{\mid b_1 =0= b_2 =b_3} = x_0. 
\end{equation}
(In what follows we use $\epsilon :=b_3$ to simplify notation.)

Acting of $\partial/\partial\epsilon $ on (B5) and making use of (B4) gives
\begin{equation}
\frac{d^2 x}{d\epsilon ^2}-6x\frac{dx}{d\epsilon}+4x^3 =0.
\end{equation}
To reduce the order of (B8) we introduce $p:=dx/d\epsilon$, which leads 
to the equation
\begin{equation}
p\frac{dp}{dx}-6xp+4x^3 =0.
\end{equation}
Eq. (B9) becomes homogeneous for $z^2 :=p$, since we get
\begin{equation}
\frac{dz}{dx}=\frac{3xz^2 -2x^3}{z^3}.
\end{equation}
Substitution $z:=ux$ into (B10) gives
\begin{equation}
\frac{u^3 du}{-u^4 +3u^2 -2}=\frac{dx}{x}.
\end{equation}
One more substitution $v:=u^2$ turns (B11) into
\begin{equation}
\Bigl(\frac{1}{v-1}-\frac{2}{v-2}\Bigr)dv = \frac{2}{x}dx.
\end{equation}
Solution to (B12) reads
\begin{equation}
\frac{v-1}{(v-1)^2} = Cx^2,
\end{equation}
where $R^1\ni C>0$ is a constant.

\noindent
Making use of of $p=dx/d\epsilon,~p=z^2,~z=ux$ and $v=u^2$ turns (B13) 
into an algebraic equation 
\begin{equation}
\Bigl(\frac{dx}{d\epsilon}\Bigr)^2 - (4x^2 +D)\frac{dx}{d\epsilon} + 4x^4 
+Dx^2 =0,
\end{equation}
where $D:=1/C.$

\noindent
Eq. (B14) splits into two first-order real equations.
One of them has the form (Analysis of the other one can be done by analogy.)
\begin{equation}
2\frac{dx}{d\epsilon}= 4x^2 +D-\sqrt{D(4x^2 +D)} .
\end{equation}
The solution to (B15) reads
\begin{equation}
\epsilon (x) = 2\int\frac{dx}{4x^2 +D-\sqrt{D(4x^2 +D)}}=\frac{1}{A-x-
\sqrt{x^2 +A^2}}+B ,
\end{equation}
where $A=\sqrt{D}/2 $ and $B$ are real constants.

\noindent
Eq. (B16) leads to
\begin{equation}
x(\epsilon)=\frac{A(\epsilon -B)\Bigl[A(\epsilon -B)-1\Bigr] +1}
{2 (\epsilon -B)\Bigl[A(\epsilon -B)-1\Bigr]} .
\end{equation}
Eq. (B17) represents one of the solutions of (B5). It is not defined for 
$\epsilon =B$ because
\begin{equation}
\lim_{\epsilon\rightarrow B-}x(\epsilon) =+\infty,~~~~
\lim_{\epsilon\rightarrow B+}x(\epsilon) =-\infty .
\end{equation}
Since (B17) is not defined for all $\epsilon \in R$ , we conclude 
that the vector field $X_3$ is not complete  on the plane.

One can easily solve the Lie equations corresponding 
to (B1) and (B2). The solutions, respectively, read
\begin{equation}
(t,~x)\longrightarrow (t,~x+b_0)
\end{equation}
and 
\begin{equation}
(t,~x)\longrightarrow (t-rb_1,~x\exp b_1).
\end{equation}
Both (B19) and (B20) describe one-parameter global transformations on $V_p$ 
well defined for any $b_0,b_1 \in \dR$. Therefore, the vector fields $X_1$ 
and $X_2$ are complete on the plane.

\section{Representation algebra on plane}

In what follows we consider only the case $\hat{A}\equiv \hat{A}_-$ and 
$\hat{B}\equiv \hat{B}_-$ (another case can be done by analogy).

We give the proof that representation of the algebra
\begin{equation}
\{A,B\}=B
\end{equation}
defined by 
\begin{equation}
\hat{B}\phi (x):= x\phi(x),~~~~\hat{A}\phi(x) := -i x \frac{d}{dx}\phi(x) 
~~~~x\in \dR_+,~~~~\phi \in \Lambda =C^\infty_0(\dR_+) \subset 
\mathcal{H}_\Lambda
\end{equation}
with
\begin{equation}
<\phi _1| \phi _2>= \int^\infty_0~\overline{\phi_1(x)}\phi_2(x)\frac{dx}{x},
~~~~\phi_1,\phi_2 \in \Lambda
\end{equation}
is essentially self-adjoint on $\Lambda$ (the space $\mathcal{H}_{\Lambda}$ 
denotes the 
completion of $\Lambda$ with respect to the inner product (C3)):

\noindent 
It is easy to see that the representation (C2) and is symmetric on 
a common invariant dense domain $\Lambda$.

\noindent
To examine the self-adjointness of $\hat{A}$ we solve the equation 
\begin{equation}
\hat{A}^\ast f_\pm(x)=\pm if_\pm(x),~~~~f_\pm \in D(\hat{A^\ast})
\subset \mathcal{H}_\Lambda
\end{equation}
to find the deficiency indices $n_+(\hat{A})$ and $n_-(\hat{A})$.
The solution to (C4) reads 
\begin{equation}
f_\pm(x)=a_\pm x^{\pm 1},
\end{equation}
where $a_\pm \in \dC$. 

\noindent
It is clear that $f_\pm $ are not in $\mathcal{H}_\Lambda$ unless $a_\pm =0$. 
Thus $n_+(\hat{A})=0=n_-(\hat{A})$, which means \cite{MRS} that $\hat{A}$ 
is essentially self-adjoint on $\Lambda$.

The case of $\hat{B}$ operator is trivial since
\begin{equation}
\hat{B}^\ast g_\pm(x) = \pm ig_\pm(x),~~~~g_\pm \in D(\hat{B}^\ast)
\subset \mathcal{H}_\Lambda
\end{equation}
reads $(x\mp i) g_\pm(x)=0.$ Its only solutions are $g_\pm(0)=0$, which proves 
that $n_+(\hat{B})=0=n_-(\hat{B})$.
Therefore, Eqs. (C2) and (C3) define an essentially self-adjoint 
representation of (C1) algebra. 

\begin{acknowledgments}
I am  very grateful to J-P. Gazeau, G. Jorjadze, S. Woronowicz  and  
A. Trautman for valuable discussions, and to the anonymous referee for 
constructive criticism.
 
\end{acknowledgments}

\end{document}